\documentclass[11pt]{article}
\usepackage{amsmath} % for typesetting matrices
\usepackage{amssymb}
\usepackage{bm} % bold math
\usepackage{graphicx}
\usepackage[colorlinks=true,citecolor=blue,linkcolor=black]{hyperref}
\numberwithin{equation}{section}

\def\x{{\bf x}}
\def\k{{\bf k}}

\def\v{{\bf v}}

\def\coeff#1#2{{\textstyle {\frac {#1}{#2}}}}

\begin{document}

\title{\Large Stable and causal relativistic Navier-Stokes equations}
\author{\normalsize Raphael E. Hoult and Pavel Kovtun\\
{\it\normalsize Department of Physics \& Astronomy,  University of Victoria,}\\
{\it\normalsize PO Box 1700 STN CSC, Victoria, BC,  V8W 2Y2, Canada}
}

\date{\normalsize April, 2020}

\maketitle

\begin{abstract}
\noindent
Relativistic Navier-Stokes equations express the conservation of the energy-momentum tensor and the particle number current in terms of the local hydrodynamic variables: temperature, fluid velocity, and the chemical potential. We show that the viscous-fluid equations are stable and causal if one adopts suitable non-equilibrium definitions of the hydrodynamic variables. 
\end{abstract}

\section{Introduction}
Hydrodynamics is a classical effective description of macroscopic states of matter with small deviations from local thermal equilibrium. Hydrodynamics is conventionally formulated by starting from thermodynamics, and then promoting the constant parameters of global thermal equilibrium (temperature $T$, fluid velocity $\v$, etc.) to slowly varying functions in space and time: $T(t,\x)$, $\v(t,\x)$, etc. The evolution of these hydrodynamic variables is then determined by the local conservation laws of energy, momentum, and possibly other conserved quantities such as mass or particle number~\cite{LL6}. Intrinsic to hydrodynamics is thus a fundamental ambiguity: one must (somewhat arbitrarily) make a choice as to how to define ``local temperature'', ``local fluid velocity'', etc., out of equilibrium. For example, the non-equilibrium ``fluid velocity'' may be chosen to correspond to the flow of particles, or to the flow of energy, or to the flow of entropy, etc., with each choice resulting in different hydrodynamic equations. 

In non-relativistic Navier-Stokes equations, the standard convention is to define the ``fluid velocity'' through the flow of mass~\cite{LL6}. Relativistic hydrodynamics was presented originally in two different formulations, one pioneered by Eckart~\cite{PhysRev.58.919} and one by Landau and Lifshitz~\cite{LL6}; both are still widely discussed. The non-equilibrium conventions differ between the two formulations. Consequently, the hydrodynamic equations of Eckart and of Landau and Lifshitz are different, mathematically inequivalent, equations.
More generally, the arbitrariness in adopting different non-equilibrium definitions implies that there is simply no such thing as ``the'' equations of hydrodynamics. Still, one expects that conventions should not be physically relevant, and that different hydrodynamic equations must give rise to the same physical predictions within the domain of applicability of hydrodynamics.

The standard formulation of relativistic hydrodynamics (as in Ref.~\cite{LL6}) uses the following variables: temperature $T$, chemical potential $\mu$, and fluid velocity $u^\alpha$. The first two are scalars, and the latter is a vector. The chemical potential is conjugate to a global conserved $U(1)$ charge, such as the baryon number. The hydrodynamic equations are the conservation laws for the energy-momentum tensor $T^{\alpha\beta}$ and the corresponding $U(1)$ current $J^\alpha$,
\begin{align}
\label{eq:TJcons}
  \nabla_{\alpha}T^{\alpha\beta} = 0\,, \ \ \ \ \nabla_\alpha J^\alpha = 0\,,
\end{align}
where $\nabla$ denotes the covariant derivative for the spacetime metric $g_{\alpha\beta}$. The conservation laws have to be supplemented with the constitutive relations which express $T^{\alpha\beta}$ and $J^\alpha$ in terms of $T$, $\mu$, and $u^\lambda$. We will take the constitutive relations $T^{\alpha\beta} = T^{\alpha\beta}[T,\mu, u^\lambda, g_{\rho\sigma}]$, $J^{\alpha} = J^{\alpha}[T,\mu, u^\lambda, g_{\rho\sigma}]$ to be local functions of the hydrodynamic variables, the metric, and their derivatives. The constitutive relations are then written as an expansion in derivatives,
\begin{align}
  & T^{\alpha\beta} = T^{\alpha\beta}_{(0)}[T,\mu, u, g] + T^{\alpha\beta}_{(1)}[\partial T,\partial \mu, \partial u, \partial g] + \dots \,,\\
  & J^{\alpha} = J^{\alpha}_{(0)}[T,\mu, u, g] + J^{\alpha}_{(1)}[\partial T,\partial \mu, \partial u, \partial g] + \dots \,,
\end{align}
where the subscript denotes the number of derivatives. For example, $T^{\alpha\beta}_{(2)}$ will have contributions proportional to $(\partial T)(\partial\mu)$, $\partial^2 u$, as well as the purely geometric contributions proportional to the Ricci tensor $R^{\alpha\beta}$, among others. The constitutive relations which only take into account $ T^{\alpha\beta}_{(0)}$, $ J^{\alpha}_{(0)}$ are said to correspond to ``perfect fluid hydrodynamics'', while the constitutive relations with terms up to $T^{\alpha\beta}_{(n)}$, $J^\alpha_{(n)}$ are said to describe ``$n$-th order hydrodynamics''. The standard physics of viscosity and heat conductivity is contained within first-order hydrodynamics. The equations of first-order hydrodynamics are often called the Navier-Stokes equations.

The first-order relativistic theories of Eckart, and of Landau and Lifshitz, suffer from two important pathologies: they both predict that the uniform thermal equilibrium state of a non-gravitating fluid in flat space is unstable~\cite{Hiscock:1985zz}, and they both predict that signals propagate faster than light~\cite{Hiscock:1987zz}. The most popular remedy to these problems is provided by the M\"uller-Israel-Stewart (MIS) theories which introduce extra tensor variables besides $T$, $u^\alpha$, and $\mu$ into the hydrodynamic equations. See the recent book~\cite{Romatschke:2017ejr} for a modern perspective and references. The pathologies of the original hydrodynamic theories of \cite{LL6, PhysRev.58.919} plus the successful practical applications of the MIS theories in describing the quark-gluon plasma produced in relativistic heavy-ion collisions~\cite{Gale:2013da, Jeon:2015dfa} have led to a widespread belief which can be stated as ``the relativistic Navier-Stokes equations are unstable and acausal''.

Such a view, however, is misguided: as previously emphasized, there is no such thing as ``the'' relativistic Navier-Stokes equations. The freedom of convention%
\footnote{ 
  Different conventions correspond to arbitrariness in performing derivative field redefinitions of the hydrodynamic variables. In the literature on relativistic hydrodynamics, such redefinitions are often referred to as different ``frames''. See e.g.\ Ref.~\cite{Kovtun:2019hdm} for a discussion of different conventions and how to translate between them.
}
in defining the out-of-equilibrium $T$, $u^\alpha$, and $\mu$ means that for the same physical fluid there are infinitely many different ``Navier-Stokes equations''; the theories of~\cite{LL6} and \cite{PhysRev.58.919} are just two examples. With some conventions, the Navier-Stokes equations are indeed unstable and acausal; with others, the Navier-Stokes equations may well  be both stable and causal. One may as well choose a convention that makes physical sense. 

In Refs.~\cite{Kovtun:2019hdm, Bemfica:2017wps, Bemfica:2019knx} (which we shall call BDNK) it was argued that there exist particularly convenient out-of-equilibrium definitions of the hydrodynamic variables, such that the equations of first-order hydrodynamics written in terms of these variables are causal, and the equilibrium state is stable. The BDNK conventions thus define stable and causal frames for relativistic fluids. The discussion in Refs.~\cite{Bemfica:2017wps, Bemfica:2019knx} was only concerned with ``uncharged'' fluids, i.e.\ fluids for which the only hydrodynamic variables are $T$ and $u^\alpha$, and the only conservation laws are those of energy and momentum. In Ref.~\cite{Kovtun:2019hdm} general hydrodynamic field redefinitions were discussed for fluids with a global $U(1)$ charge such as the baryon number (so called ``charged'' fluids); however, the stable and causal frames were only explicitly discussed for uncharged fluids. 

The aim of the present paper is to extend the discussion of BDNK to fluids with a conserved $U(1)$ charge, whose hydrodynamic equations are given by the conservation laws~(\ref{eq:TJcons}). For such fluids, we shall describe a class of stable and causal frames. In such frames, the relativistic Navier-Stokes equations are stable and causal, correcting the deficiencies of \cite{LL6, PhysRev.58.919}, but without introducing any extra variables besides the standard $T$, $\mu$, and $u^\alpha$. The relation between the familiar Landau-Lifshitz frame and the general frames is described in the Appendix.

\section{Constitutive relations}
Following Ref.~\cite{PhysRev.58.919}, we decompose the energy-momentum tensor and the current as
\begin{subequations}
\label{eq:TJ1}
\begin{align}
  &  T^{\mu\nu} = {\cal E} u^\mu u^\nu + {\cal P} \Delta^{\mu\nu} +
     ({\cal Q}^\mu u^\nu + {\cal Q}^\nu u^\mu) + {\cal T}^{\mu\nu}\,,\\
  & J^\mu = {\cal N} u^\mu + {\cal J}^\mu\,,
\end{align}
\end{subequations}
where $\Delta^{\mu\nu}\equiv g^{\mu\nu} + u^\mu u^\nu$ projects onto the space orthogonal to $u$, the vectors ${\cal Q}$ and ${\cal J}$ are transverse to $u$, and the tensor ${\cal T}$ is transverse to $u$, symmetric and traceless. For a given timelike vector $u$, the decomposition (\ref{eq:TJ1}) defines the components ${\cal E}$, ${\cal Q}^\mu$, ${\cal T}^{\mu\nu}$, ${\cal N}$, and ${\cal J}^\mu$ in terms of $T^{\mu\nu}$ and $J^\mu$. 
Using the notation of~\cite{Kovtun:2019hdm}, the most general one-derivative
constitutive relations in a charged fluid may be written as
\begin{subequations}
\label{eq:EPQTNJ1}
\begin{align}
 {\cal E} & = \epsilon + \varepsilon_1 \dot T/T + \varepsilon_2 \nabla_{\!\lambda} u^\lambda + \varepsilon_3 u^\lambda \partial_\lambda(\mu/T) + O(\partial^2)\,, \\[5pt]
 {\cal P} & = p + \pi_1 \dot T/T + \pi_2 \nabla_{\!\lambda} u^\lambda +\pi_3 u^\lambda \partial_\lambda(\mu/T) + O(\partial^2)\,, \\[5pt]
 {\cal Q}^\mu & = \theta_1  \dot u^\mu + \theta_2/T \,\Delta^{\mu\lambda} \partial_\lambda T + \theta_3 \Delta^{\mu\lambda} \partial_\lambda (\mu/T) +  O(\partial^2)\,,\\[5pt]
 {\cal T}^{\mu\nu} & = -\eta \sigma^{\mu\nu} + O(\partial^2)\,,\\[5pt]
 {\cal N} & = n + \nu_1 \dot T/T + \nu_2 \nabla_{\!\lambda} u^\lambda + \nu_3 u^\lambda \partial_\lambda(\mu/T) + O(\partial^2)\,, \\[5pt]
 {\cal J}^\mu & = \gamma_1  \dot u^\mu + \gamma_2/T \,\Delta^{\mu\lambda} \partial_\lambda T + \gamma_3 \Delta^{\mu\lambda} \partial_\lambda (\mu/T) +  O(\partial^2)\,.
\end{align}
\end{subequations}
The dot signifies the derivative along the fluid velocity, i.e.\ $\dot T\equiv u^\lambda \partial_\lambda T$, $\dot u^\alpha \equiv u^\lambda \nabla_{\!\lambda} u^\alpha$. The shear tensor is
$$
  \sigma^{\mu\nu} = \Delta^{\mu\rho}\Delta^{\nu\sigma} \left( \nabla_{\!\rho} u_\sigma + \nabla_\sigma u_\rho -\coeff{2}{d}\, g_{\rho\sigma} \nabla_{\!\alpha}u^\alpha \right)\,,
$$
where $d$ is the number of spatial dimensions. We will refer to the coefficients $\varepsilon_i$, $\pi_i$, $\theta_i$, $\nu_i$, $\gamma_i$ as ``transport parameters''. There are fifteen transport parameters at one-derivative order, plus the shear viscosity $\eta$. We will work in the thermodynamic frame where the equilibrium constitutive relations follow from a partition function which is extensive in equilibrium.%
\footnote{
  The equilibrium partition function is a functional of the external time-independent sources: the metric $g_{\mu\nu}$ and the gauge field $A_\mu$ that couples to the $U(1)$ current. Equilibrium fluid velocity is defined to be aligned with the timelike Killing vector field $V$ which specifies the direction of time, $u^\mu = V^\mu/\sqrt{-V^2}$, to all orders in the derivative expansion. Equilibrium temperature is defined so that the Tolman's law $T=T_0/\sqrt{-V^2}$ holds, to all orders in the derivative expansion. See Ref.~\cite{Jensen:2012jh} for more details. Comparing $T^{\mu\nu}$ and $J^\mu$ obtained by varying the equilibrium partition function with respect to $g_{\mu\nu}$ and $A_\mu$ with $T^{\mu\nu}$ and $J^\mu$ given by the constitutive relations (\ref{eq:TJ1}), (\ref{eq:EPQTNJ1}) and evaluated in equilibrium, one obtains the constraints stated in the text.
}
Then the zero-derivative coefficients $\epsilon(T,\mu)$, $p(T,\mu)$, $n(T,\mu)$ have the standard interpretations of the equilibrium energy density, pressure, and charge density, respectively, and moreover are related by $n = \partial p/\partial\mu$, $\epsilon + p -\mu n = T \frac{\partial p}{\partial T}$. Further, the same extensivity in the thermodynamic frame implies $\theta_1=\theta_2$, and $\gamma_1 = \gamma_2$.

The bulk viscosity $\zeta$ and the charge conductivity $\sigma$ are given by the following combinations of the transport parameters~\cite{Kovtun:2019hdm}:
\begin{align}
\label{eq:zeta-inv}
  & \zeta = (p_{,\epsilon}\pi_1 - \pi_2) + p_{,\epsilon}(\varepsilon_2 - p_{,\epsilon}\varepsilon_1) +  \frac{p_{,n}}{T}(\pi_3 - p_{,\epsilon} \varepsilon_3) + p_{,n}(\nu_2 - p_{,\epsilon} \nu_1)  - \frac{p_{,n}^2}{T} \nu_3 \,,\\
\label{eq:sigma-inv}
  & \sigma = -\frac{\gamma_3}{T} + \frac{n(T\gamma_1+ \theta_3)}{T(\epsilon+p)} - \frac{n^2 \theta_1}{(\epsilon+p)^2} \,,
\end{align}
where the derivatives of the pressure are $p_{,\epsilon}\equiv (\partial p/\partial\epsilon)_n$, $p_{,n}\equiv (\partial p/\partial n)_{\epsilon}$. Clearly, the transport parameters must be such that both $\zeta$ and $\sigma$ are non-negative. A choice of ``frame'' corresponds to a choice of transport parameters. For example, the Landau-Lifshitz frame of~\cite{LL6} imposes that all one-derivative transport parameters vanish except $\pi_2$ and $\gamma_3$. A stable and causal frame is a choice of the transport parameters such that the hydrodynamic equations (\ref{eq:TJ1}), (\ref{eq:EPQTNJ1}) are causal and predict that the thermal equilibrium state is stable.

In the grand canonical ensemble, the functions $\epsilon(T,\mu)$ and $n(T,\mu)$ are not independent, as both are determined by the pressure $p(T,\mu)$. In particular, $T (\partial n/\partial T) + \mu (\partial n/\partial\mu) = \partial\epsilon/\partial\mu$.
Further, we have the following thermodynamic inequalities:
\begin{align}
\label{eq:thermo-ineq}
  \frac{\partial n}{\partial\mu}\geqslant 0\,,\ \ \ \ 
  T \frac{\partial\epsilon}{\partial T} + \mu \frac{\partial\epsilon}{\partial\mu} \geqslant 0\,,\ \ \ \ 
  \frac{\partial\epsilon}{\partial T} \frac{\partial n}{\partial\mu} - \frac{\partial n}{\partial T} \frac{\partial\epsilon}{\partial\mu} \geqslant 0\,.
\end{align}
Denoting the Hamiltonian by $H$ and the conserved particle number operator by $N$, the inequalities above follow by demanding that the connected equilibrium functions are non-negative: $\langle N^2\rangle_{\rm conn} \geqslant 0$, $\langle H^2\rangle_{\rm conn} \geqslant 0$, $\langle H^2\rangle_{\rm conn} \langle N^2\rangle_{\rm conn} - \langle HN\rangle_{\rm conn}^2 \geqslant 0$. 
The inequalities (\ref{eq:thermo-ineq}) imply
\begin{align}
\label{eq:ineq-2}
  p_{,\epsilon} + \frac{(\partial n/\partial\mu)}{\epsilon+p}p_{,n}^2 \geqslant \frac{n}{\epsilon+p} p_{,n} \,,\ \ \ \
  \frac{\partial n}{\partial\mu} \geqslant \frac{n^2}{(\epsilon+p) v_s^2}\,,
\end{align}
where $v_s$ is the speed of sound, see e.g.~\cite{Kovtun:2012rj}.

The general constitutive relations (\ref{eq:EPQTNJ1}) simplify if the underlying microscopic theory happens to be conformal. In $d+1$ spacetime dimensions, conformal symmetry demands~\cite{Kovtun:2019hdm}:
\begin{align}
\label{eq:CFTc}
  \epsilon = d p\,,\ \ \ \ \varepsilon_i = d \pi_i\,, \ \ \ \ \pi_1 = d \pi_2\,, \ \ \ \ \nu_1 = d \nu_2\,,
\end{align}
as well as $\theta_1 = \theta_2$, $\gamma_1 = \gamma_2$. The equation of state in a conformal theory is $p(T,\mu) = T^{d+1} f(\mu/T)$, with a dimensionless function $f(\mu/T)$ which is determined by the microscopic dynamics. The inequalities (\ref{eq:thermo-ineq}) imply that the function $f(x)$ must be such that 
\begin{align}
  f'' \geqslant 0\,,\ \ \ \ f f'' \geqslant \frac{d}{d+1} (f')^2 \,.
\end{align}
The speed of sound in a conformal theory is $v_s = 1/\sqrt{d}$, and the bulk viscosity $\zeta$ vanishes.

\section{Small fluctuations in equilibrium}
\label{sec:linear-analysis}
Let us now look at small fluctuations of the equilibrium state with constant $T=T_0$, $\mu=\mu_0$, and $\v=\v_0$. Taking the fluctuations $\delta T$, $\delta\mu$, $\delta v_i$ proportional to $e^{-i\omega t+ i\k{\cdot}\x}$, the linearized hydrodynamic equations give rise to polynomial equations in $\omega$ and $\k$ which we schematically write as $F(\omega,\k)=0$. Their solutions determine the dispersion relations $\omega = \omega_a(\k)$. We take $\k$ to be real; the $\omega_a(\k)$ will be (in general, complex) functions of $k\equiv|\k|$ and $(\k{\cdot}\v_0)$. The modes with $\omega_a(\k{\to}0)=0$ are ``gapless'', while the modes with $\omega_a(\k{\to}0)\neq0$ are ``gapped''. All genuine hydrodynamic modes (sound waves, shear waves, heat diffusion) are gapless, reflecting the existence of conserved densities. On the other hand, the gapped modes, if they are present, should be viewed as parametrizations of non-hydrodynamic physics.

\subsection*{Hydrodynamic modes}
There are $d+2$ hydrodynamic (gapless) modes: $d{-}1$ transverse shear modes, two sound modes, and one heat diffusion mode. Their dispersion relations for the fluid at rest ($\v_0=0$) take the following form at small~$\k$, as described for example in Ref.~\cite{Kovtun:2012rj}:
\begin{align}
\label{eq:w-shear}
  & \omega_{\rm shear}(\k) = -\frac{i\eta}{\epsilon + p} \k^2 + \dots \,,\\
\label{eq:w-sound}
  & \omega_{\rm sound}(\k) = \pm v_s |\k| -\frac{i}{2}\Gamma_s \k^2 + \dots\,,\\
\label{eq:w-D}
  & \omega_{\rm heat}(\k) = -i D \k^2 + \dots \,.
\end{align}
The speed of sound is expressed in terms of the equilibrium thermodynamic quantities as
\begin{align}
  v_s^2 = \left(\frac{\partial p}{\partial\epsilon}\right)_{\!n} + \frac{n}{\epsilon+p}\left( \frac{\partial p}{\partial n}\right)_{\!\epsilon} \,,
\end{align}
which can also be written as $v_s^2 = (\partial p/\partial\epsilon)_S$. 
The damping coefficient of the sound waves is
\begin{align}
  \Gamma_s = \frac{(2-\coeff{2}{d}) \eta+\zeta}{\epsilon + p} + \frac{\sigma}{(\epsilon{+}p) v_s^2}\left( \frac{\partial p}{\partial n} \right)_{\!\epsilon}^2\,,
\end{align}
determined by the viscosities and the charge conductivity.
The heat diffusion coefficient is
\begin{align}
  D = \frac{\sigma(\epsilon{+}p) (\partial p/\partial\epsilon)_n^2}{v_s^4(\epsilon{+}p)(\partial n/\partial\mu) - n^2 v_s^2}\,.
\end{align}
Thermodynamic inequality (\ref{eq:ineq-2}) implies that $D$ is positive for positive $\sigma$. For the state with $n=0$ in equilibrium, the diffusion constant is related to the conductivity by $\sigma = (\partial n/\partial\mu) D$. 

The hydrodynamic (gapless) dispersion relations in a moving fluid with $\v_0\neq0$ can be found by applying a Lorentz boost to the spectral function $F(\omega,\k, \v_0{=}0)$, as described for example in Ref.~\cite{Kovtun:2019hdm}. If a mode has a quadratic dispersion relation $\omega = -iD\k^2 + \dots$ at small $\k$ in the fluid at rest, then in a moving fluid one finds at small $\k$:
\begin{align}
  \omega = \k{\cdot}\v_0 - iD\sqrt{1{-}\v_0^2}\left( \k^2 - (\k{\cdot}\v_0)^2\right) + \dots \,.
\end{align}
The corresponding formulas for the sound mode at $\v_0\neq0$ can be found in Ref.~\cite{Kovtun:2019hdm}.

\subsection*{Stability and causality}

We will call the $a$-th mode ``stable'' if
\begin{align}
\label{eq:stability}
  {\rm Im}\; \omega_a(\k) \leqslant 0\,,
\end{align}
and we will call the $a$-th mode ``causal'' if
\begin{align}
\label{eq:causality}
  0< \lim_{k\to\infty} \left| \frac{{\rm Re}\; \omega_a(\k)}{k} \right| < 1\,.
\end{align}
See Ref.~\cite{Fox:1970cu} for a discussion of causality and large-$k$ dispersion relations.
The modes are as follows. The shear-channel fluctuations decouple from the sound-channel fluctuations as a consequence of rotation invariance, so that $F(\omega,\k) = F_{\rm shear}(\omega,\k)^{d-1} F_{\rm sound}(\omega,\k)$. The function $F_{\rm shear}(\omega,\k)$ is a second-order polynomial in $\omega$, which gives rise to one gapless mode (\ref{eq:w-shear}) and one gapped mode. The function $F_{\rm sound}(\omega,\k)$ is a sixth-order polynomial in $\omega$ which gives rise to three gapless modes (\ref{eq:w-sound}), (\ref{eq:w-D}) and three gapped modes. 

For shear-channel fluctuations, the modes of a charged fluid are identical to those of an uncharged fluid. Demanding stability and causality then gives rise to the constraint~\cite{Kovtun:2019hdm}
\begin{align}
\label{eq:shear-constraint}
  \theta_1 > \eta >0 \,.
\end{align}
Let us now look at sound-channel fluctuations. The spectral function $F_{\rm sound}(\omega,\k)$ is lengthy, and while one could in principle derive the  constraints by applying Eqs.~(\ref{eq:stability}), (\ref{eq:causality}) to the roots of $F_{\rm sound}(\omega,\k) = 0$, the constraints are unwieldy, and depend on the equation of state. The spectral function can be analyzed in various limiting cases in order to derive various necessary or sufficient conditions for stability and causality. 

One set of necessary conditions follows by requiring that the gaps (obtained by solving $F_{\rm sound}(\omega,\k{=}0) = 0$) have negative imaginary parts. Demanding that the sound-channel gaps are stable at $\v_0=0$ gives the following necessary conditions for the stability of equilibrium:
\begin{align}
\label{eq:stabgap-1}
  & \varepsilon_1 \nu_3 - \varepsilon_3 \nu_1 > 0\,,\\
\label{eq:stabgap-2}
  & \nu_3 \left(\frac{\partial\epsilon}{\partial T}\right)_{\!\mu/T}  + \varepsilon_1 \left(\frac{\partial n}{\partial\mu} \right)_{\!T} - \nu_1 \left(\frac{\partial \epsilon}{\partial\mu} \right)_{\!T} - \varepsilon_3 \left(\frac{\partial n}{\partial T}\right)_{\!\mu/T}  >0\,.
\end{align}
Note that $T(\partial\epsilon/\partial T)_{\mu/T} = T(\partial\epsilon/\partial T) + \mu(\partial\epsilon/\partial\mu)\geqslant0$, thanks to the thermodynamic inequalities (\ref{eq:thermo-ineq}). For example, in a frame with $\varepsilon_3 = \nu_1 = 0$, stability conditions (\ref{eq:stabgap-1}), (\ref{eq:stabgap-2}) will be satisfied for $\varepsilon_1 >0$, $\nu_3>0$.

The constraints on the transport parameters arising from the stability of the gaps in a moving fluid are more involved. With $\Delta\equiv-i\omega$, the gaps satisfy a cubic equation $a_0 \Delta^3 + a_1 \Delta^2 + a_2 \Delta + a_3 = 0$, with coefficients $a_n$ that depend on the transport parameters and on the equation of state. The coefficient $a_3$ is proportional to $(\frac{\partial\epsilon}{\partial T} \frac{\partial n}{\partial\mu} {-} \frac{\partial n}{\partial T} \frac{\partial\epsilon}{\partial\mu})(1{-}v_0^2 v_s^2)$, hence one can always take $a_3>0$. Demanding that the sound-channel gaps are stable, i.e.\ ${\rm Re}\,\Delta <0$, by the Routh-Hurwitz criterion~\cite{Korn-Korn} then amounts to $a_0>0$, $a_1>0$, $a_1 a_2 > a_0 a_3$.

Another simple set of constraints comes from the high-momentum modes. As $k\to\infty$, the six sound-channel modes have linear dispersion relations $\omega = \pm c_s k$, where $c_s^2$ satisfies a cubic equation. The transport coefficients must be constrained by demanding $c_s^2>0$ (stability) and $c_s^2<1$ (causality). If we choose a frame with $\varepsilon_3 = \pi_3 = \theta_3 = 0$, the cubic equation for $c_s^2$ factorizes into a product of a linear equation and a quadratic equation: 
\begin{align}
\label{eq:cc-1}
  & \nu_3 c_s^2 + \gamma_3 = 0\,,\\
\label{eq:cc-2}
  & \varepsilon_1 \theta_1 (c_s^2)^2  - c_s^2 (\theta_1 \pi_1 + \varepsilon_2 (\theta_1 {+} \pi_1) + \varepsilon_1 (\coeff{2d-2}{d} \eta {-} \pi_2))  - \theta_1 (\coeff{2d-2}{d} \eta{-}\pi_2) = 0 \,.
\end{align}
The first one implies 
\begin{align}
\label{eq:cc-g}
  0 < -\gamma_3/\nu_3 < 1\,.
\end{align}
The constraints on the transport coefficients from Eq.~(\ref{eq:cc-2}) can be obtained as follows. For a quadratic equation $ax^2 + bx+c = 0$ with $a>0$, the conditions that the roots are real and fall between 0 and 1 amount to the following: 
\begin{align}
\label{eq:cc-3}
  b^2 - 4ac>0\,,\ \ \ \ b<0\,,\ \ \ \ 0<c<a\,,\ \ \ \ a+b+c>0\,.
\end{align}
Applying these to Eq.~(\ref{eq:cc-2}) with 
\begin{align}
\label{eq:abc-1}
   a=\varepsilon_1 \theta_1\,,\ \ \ \  
   b=-\theta_1 \pi_1 - \varepsilon_2 (\theta_1 {+} \pi_1) - \varepsilon_1 (\coeff{2d-2}{d} \eta{-}\pi_2)\,,\ \ \ \ 
   c = \theta_1 (\pi_2{-}\coeff{2d-2}{d} \eta) 
\end{align}
gives a set of non-linear constraints among the coefficients $\varepsilon_{1,2}$, $\pi_{1,2}$, $\theta_1$, and $\eta$. Note that Eq.~(\ref{eq:cc-2}) is exactly the same equation that determines the propagation speed of the large-$k$ eigenmodes in an uncharged fluid~\cite{Kovtun:2019hdm}. Thus in the frame with $\varepsilon_3 = \pi_3 = \theta_3 = 0$, the causality constraints (\ref{eq:cc-3}), (\ref{eq:abc-1}) on the coefficients $\varepsilon_{1,2}$, $\pi_{1,2}$, $\theta_1$, and $\eta$ will be exactly the same as in uncharged fluids. In order to express the causality constraints in terms of physical transport coefficients, we need the bulk viscosity (\ref{eq:zeta-inv}) and the conductivity (\ref{eq:sigma-inv}) in the frame with $\varepsilon_3 = \pi_3 = \theta_3 = 0$.
From Eq.~(\ref{eq:cc-g}) one immediately finds
\begin{align}
\label{eq:nu3-constraint}
  \nu_3 > \sigma T + \frac{n^2 T\theta_1}{(\epsilon+p)^2} - \frac{n T \gamma_1}{\epsilon+p}\,.
\end{align}
For example, if we choose a frame in which $\gamma_1 = n\theta_1/(\epsilon{+}p)$, then $\gamma_3 = -\sigma T$, and the causality constraint (\ref{eq:cc-g}) becomes simply $\nu_3 > \sigma T$. Similarly, in order to simplify the bulk viscosity one could further choose a frame in which $\nu_2 = (p_{,\epsilon})\nu_1 + (p_{,n})\nu_3/T$. Then the transport parameters $\nu_i$ drop out from the bulk viscosity, and the large-$k$ causality constraints (\ref{eq:cc-3}), (\ref{eq:abc-1}) take exactly the same form as the corresponding constraints in uncharged fluids~\cite{Kovtun:2019hdm}. Expressed in terms of $\gamma_s \equiv \frac{2d-2}{d} \eta + \zeta$, the large-$k$ causality constraints (\ref{eq:cc-g}), (\ref{eq:cc-3}), (\ref{eq:abc-1}) in this frame  become: 
\begin{align}
\label{eq:cc-sound-1}
   & \nu_3 > \sigma T \,,\\
\label{eq:cc-sound-2}
   & 0 < p_{,\epsilon} (\varepsilon_2 {+} \pi_1) -(p_{,\epsilon})^2 \varepsilon_1 - \gamma_s < \varepsilon_1 \,, \\
\label{eq:cc-sound-3}
   & \varepsilon_1 \! \left( p_{,\epsilon} (\varepsilon_2 {+} \pi_1) -(p_{,\epsilon})^2 \varepsilon_1 - \gamma_s \right) < \theta_1 (\varepsilon_2 {+} \pi_1) + \varepsilon_2 \pi_1 \,, \\
\label{eq:cc-sound-4}
   & (\varepsilon_1 {+} \theta_1)  \left( p_{,\epsilon} (\varepsilon_2 {+} \pi_1) -(p_{,\epsilon})^2 \varepsilon_1 - \gamma_s \right) + \varepsilon_1 \theta_1 > \theta_1 (\varepsilon_2 {+}\pi_1) + \varepsilon_2 \pi_1 \,, \\
\label{eq:cc-sound-5}
   & \left[ \varepsilon_1 \! \left( p_{,\epsilon} (\varepsilon_2 {+} \pi_1) {-} (p_{,\epsilon})^2 \varepsilon_1 {-} \gamma_s \right) - \theta_1 (\varepsilon_2 {+} \pi_1) - \varepsilon_2 \pi_1 \right]^2 > 4\varepsilon_1 \theta_1^2 (p_{,\epsilon} (\varepsilon_2 {+} \pi_1) {-} (p_{,\epsilon})^2 \varepsilon_1 {-} \gamma_s).
\end{align}
These causality constraints can be simultaneously satisfied~\cite{Kovtun:2019hdm}.

\subsection*{Stable and causal frames for conformal fluids}
The discussion above simplifies for conformal fluids, where the one-derivative transport parameters satisfy the constraints of Eq.~(\ref{eq:CFTc}) due to conformal symmetry. Let us further choose a frame in which $\pi_3$, $\theta_3$, $\nu_1$, and $\gamma_1$ all vanish, leaving one with five independent transport parameters $\pi_2$, $\theta_1$, $\eta$, $\nu_3$, and $\gamma_3$. 
The five parameters can be thought of as two genuine transport coefficients $\eta$ and $\sigma$, and three ``relaxation times'' corresponding to the relaxation of the energy density ($\pi_1$), momentum density ($\theta_1$), and charge density ($\nu_3$).
 
As $k\to0$, the stability conditions (\ref{eq:stabgap-1}), (\ref{eq:stabgap-2}) for the gapped modes become
$\nu_3 \pi_2 >0$, and $T\nu_3 + d \lambda \pi_2 >0$ where $\lambda\equiv T^2 (\partial n/\partial\mu)/(\epsilon{+}p) >0$ is the dimensionless charge susceptibility. These are satisfied for $\nu_3>0$, $\pi_2>0$. The large-$k$ constraint (\ref{eq:nu3-constraint}) becomes 
\begin{align}
\label{eq:CFT-constraints-1}
  T\nu_3 > T^2\sigma + \kappa^2 \theta_1\,,
\end{align}
where $\kappa\equiv nT/(\epsilon{+}p)$ is a dimensionless measure of the equilibrium charge density. Note that the thermodynamic inequalities (\ref{eq:thermo-ineq}) imply $\lambda\geqslant d\kappa^2$. In fact, the only information about the equation of state that is relevant for the linearized analysis of stability and causality is contained in $\kappa$ and $\lambda$. The large-$k$ constraints (\ref{eq:cc-sound-2}) -- (\ref{eq:cc-sound-5}) reduce simply to
\begin{align}
\label{eq:CFT-constraints-2}
  \pi_2 > \frac{2d-2}{d} \eta\,,\ \ \ \ 
  1-\frac{2d}{d{-}1}\frac{\eta}{\theta_1} - \frac{2}{d(d{-}1)}\frac{\eta}{\pi_2} >0\,.
\end{align}
Restricting to $d=3$ space dimensions, the inequalities (\ref{eq:CFT-constraints-2}) become exactly the same constraints found earlier for uncharged conformal fluids in Refs.~\cite{Bemfica:2017wps, Kovtun:2019hdm}. In order to satisfy them, it is sufficient to demand $\pi_2>\frac43\eta$, $\theta_1>4\eta$. It is a straightforward exercise to check that the large-$k$ constraints (\ref{eq:CFT-constraints-1}), (\ref{eq:CFT-constraints-2}) in $d=3$ also ensure that the sound-channel gaps are stable at all $v_0^2<1$, as long as the equation of state obeys the standard thermodynamic inequalities~(\ref{eq:thermo-ineq}). In other words, conditions (\ref{eq:CFT-constraints-1}), (\ref{eq:CFT-constraints-2}) ensure that in our chosen frame all small-$k$ modes are stable, and all large-$k$ modes are stable and causal. Finally, we point out that the large-$k$ causality alone does not guarantee stability at all $k$: for example, the causality conditions (\ref{eq:causality}) allow for negaive $\pi_2$, which is ruled out by the small-$k$ stability conditions.

\section{Real-space causality}
\label{sec:real-space}
So far, we have looked at the linearized stability of the equilibrium state, and the linearized causality of the near-equilibrium perturbations in momentum space. In fact, similar to what was done in Refs.~\cite{Bemfica:2017wps, Bemfica:2019knx}, one can study the non-linear causality of charged first-order hydrodynamics in real space using the quasi-linear character of the hydrodynamic equations. The hydrodynamic conservation laws are partial differential equations for the variables $T$, $\mu$, and $u^\alpha$ which satisfies $u_\alpha u^\alpha=-1$. Rather than working with the vector $u^\alpha$ which is constrained by $u_\alpha u^\alpha=-1$, we find it more convenient to work with the unconstrained vector $\beta^\alpha = \beta u^\alpha$, where $\beta\equiv (-\beta_\alpha \beta^\alpha)^{1/2} >0$ is the inverse temperature, $\beta=1/T$. The hydrodynamic equations (\ref{eq:TJcons}) are second-order quasilinear partial differential equations for $U^A = (\beta^\alpha, \mu/T)$ that can schematically be written as 
\begin{align}
\label{eq:UPDE}
  (M^{\mu\nu})_{AB}\, \partial_\mu \partial_\nu U^B + (N^{\mu\nu})_{ABC}\, \partial_\mu U^B \partial_\nu U^C +  (P^{\mu})_{AB}\, \partial_\mu U^B = 0 \,,
\end{align}
where the indices $A,B,C$ range from $1$ to $d{+}2$ (again, $d$ is the number of spatial dimensions), and the coefficients $M^{\mu\nu}$, $N^{\mu\nu}$, $P^\mu$ depend on $U$, but not on the derivatives of $U$. We thus have $d{+}2$ differential equations for $d{+}2$ unconstrained variables.%
\footnote{
Ref.~\cite{Bemfica:2019knx} chooses to work with $u^\alpha$ which is not constrained by $u_\alpha u^\alpha=-1$, and instead considers the projected conservation laws $u_\beta \nabla_\alpha T^{\alpha\beta} = 0$, $\Delta^\mu_{\ \;\beta} \nabla_\alpha T^{\alpha\beta} = 0$ as independent equations. Even though the norm of $u^\alpha$ is not preserved under time evolution in this approach, the causality can be argued by noting that the causal structure of the projected equations is the same as the causal structure of the original equations $\nabla_\alpha T^{\alpha\beta} = 0$. We find it conceptually cleaner to work with $\beta^\mu = u^\mu/T$ whose norm is not fixed. The causality constraints one finds by projecting the energy-momentum conservation laws are the same as those in Sec.~\ref{sec:linear-analysis}.
}
The hyperbolicity of the equations and the causality of the solutions are determined by the principal part $(M^{\mu\nu})_{AB}$, see e.g.\ Ref.~\cite{Courant-Hilbert}, Ch.~VI. The characteristic surfaces $\phi(x)={\rm const.}$ are found from
\begin{align}
\label{eq:Mxi}
  \det \left[ (M^{\mu\nu})_{AB}\, \xi_\mu\, \xi_\nu \right] = 0\,,
\end{align}
where the vectors $\xi_\mu \equiv \partial_\mu\phi(x)$ are normal to the characteristic surfaces at the point $x$. For the hydrodynamic equations (\ref{eq:UPDE}) to be hyperbolic, Eq.~(\ref{eq:Mxi}) must only have non-zero real solutions $\xi_0 = \xi_0(\xi_i)$. For the equations to be causal, the surfaces swept out by these normals must lie either outside or on the lightcone $\xi_\mu \xi^\mu = 0$.

Let us work in a frame with $\varepsilon_3 = \pi_3 = \theta_3 = 0$. The conservation laws can be written as
\begin{align}
\label{eq:dT-2}
   \nabla_\sigma T^{\sigma\alpha}  = & \ T\left[ \varepsilon_1 u^\alpha u^\rho u^\sigma u_\lambda + \varepsilon_2 u^\alpha u^\rho \Delta^\sigma_\lambda + \pi_1 \Delta^{\alpha\sigma} u^\rho u_\lambda + \pi_2 \Delta^{\alpha\sigma} \Delta^\rho_\lambda \right. \nonumber\\
  & \left. + \theta_1 (u^\alpha u^\rho \Delta^\sigma_\lambda + u^\rho u^\sigma \Delta^\alpha_\lambda ) + \theta_2 (u^\alpha u_\lambda \Delta^{\sigma\rho} + u^\rho u_\lambda \Delta^{\alpha\sigma} ) \right. \nonumber\\
  & \left. -\eta (\Delta^{\alpha\rho}\Delta^\sigma_\lambda + \Delta^\alpha_\lambda \Delta^{\sigma\rho} -\coeff2d \Delta^{\alpha\sigma} \Delta^\rho_\lambda)\right] \partial_\rho \partial_\sigma \beta^\lambda + O(\partial U \partial U) + O(\partial U)\,,
\end{align}
as well as
\begin{align}
\label{eq:dJ-2}
  \nabla_\sigma J^\sigma  = & \ T \left[ \nu_1 u^\rho u^\sigma u_\lambda + \nu_2 u^\sigma \Delta^\rho_\lambda + \gamma_1 u^\rho \Delta^\sigma_\lambda + \gamma_2 \Delta^{\rho\sigma} u_\lambda \right]\partial_\rho \partial_\sigma \beta^\lambda \nonumber\\
  & + \left[\nu_3 u^\rho u^\sigma + \gamma_3 \Delta^{\rho\sigma} \right] \partial_\rho \partial_\sigma (\mu/T) + O(\partial U \partial U) + O(\partial U)\,,
\end{align}
It is then clear that the choice of frame $\varepsilon_3 = \pi_3 = \theta_3 = 0$ ensures $(M^{\mu\nu})_{\alpha 5} = 0$, where the index ``5'' stands for $d{+}2$. The other components of the principal part, i.e.\ $(M^{\mu\nu})_{\alpha\beta}$, $(M^{\mu\nu})_{5 \alpha}$, and $(M^{\mu\nu})_{55}$ can be read off by comparing Eqs.~(\ref{eq:dT-2}), (\ref{eq:dJ-2}) to the general form (\ref{eq:UPDE}); for example, $(M^{\mu\nu})_{55} = \nu_3 u^\mu u^\nu + \gamma_3 \Delta^{\mu\nu}$. Thanks to the vanishing of $(M^{\mu\nu})_{\alpha 5}$, the determinant in Eq.~(\ref{eq:Mxi}) factorizes, and the characteristic surfaces are determined by 
\begin{align}
\label{eq:Mxi-2}
  \left[ (M^{\mu\nu})_{55}\, \xi_\mu \xi_\nu \right]  \times \det \left[ (M^{\rho\sigma})_{\alpha\beta}\, \xi_\rho\, \xi_\sigma \right] = 0\,.
\end{align}
The first factor in Eq.~(\ref{eq:Mxi-2}) gives 
\begin{align}
  \nu_3 (u{\cdot}\xi)^2 + \gamma_3 (\xi{\cdot}\Delta{\cdot}\xi) = 0\,.
\end{align}
At a given point in spacetime, passing to a local coordinate system in which $u^\alpha=(1,{\bf 0})$ at that point, the solutions are $\xi_0 = \pm \sqrt{-\gamma_3/\nu_3}\, |\xi_i|$. Demanding that these are real and lie outside the lightcone gives the same constraint (\ref{eq:cc-g}) we found earlier from the linearized analysis in momentum space. In $d{+}1$ dimensions, the determinant in Eq.~(\ref{eq:Mxi-2}) can be computed by using the following easily derived identity:
\begin{align}
\label{eq:detMM}
  & \det(A u^\alpha u_\beta + B \Delta^\alpha_{\ \;\beta} + C u^\alpha \xi_\beta + D \xi^\alpha u_\beta + E \xi^\alpha \xi_\beta) \nonumber\\
  & = B^{d-1} \left[ -AB + B(C{+}D) (\xi{\cdot}u) - BE (\xi{\cdot}u)^2 + (CD{-}AE) (\xi{\cdot}\Delta{\cdot}\xi) \right] \,.
\end{align}
Applying this to $\det \left[ (M^{\rho\sigma})^\alpha_{\ \;\beta}\, \xi_\rho\, \xi_\sigma \right]$, we find $B=\theta_1 (\xi{\cdot}u)^2 - \eta (\xi{\cdot}\Delta{\cdot}\xi)$, hence there are characteristic surfaces determined by
\begin{align}
  \left[ \theta_1 (\xi{\cdot}u)^2 - \eta (\xi{\cdot}\Delta{\cdot}\xi) \right]^{d-1} = 0\,.
\end{align}
These correspond to the $(d{-}1)$ shear-channel modes of Sec.~\ref{sec:linear-analysis}. Demanding hyperbolicity and causality then gives the constraint (\ref{eq:shear-constraint}) found earlier from the linearized analysis in momentum space. The remaining characteristic surfaces are found by setting the term in the square brackets in Eq.~(\ref{eq:detMM}) to zero, with $A,B,C,D,E$ extracted from Eq.~(\ref{eq:dT-2}). We find
\begin{align}
 \theta_2 \!\left(\coeff{2d-2}{d}\eta {-} \pi_2 \right) (\xi{\cdot}\Delta{\cdot}\xi)^2 + \left[ \varepsilon_1 \! \left(\coeff{2d-2}{d}\eta {-} \pi_2 \right) + \varepsilon_2 (\theta_2 {+} \pi_1) + \theta_1 \pi_1 \right] (\xi{\cdot}\Delta{\cdot}\xi) (\xi{\cdot}u)^2 - \varepsilon_1 \theta_1 (\xi{\cdot}u)^4 = 0.
\end{align}
The conditions of hyperbolicity and causality will be satisfied for $(\xi{\cdot}u) = c_s (\xi{\cdot}\Delta{\cdot}\xi)$, with $c_s$ real and $0<c_s^2<1$. This gives exactly the same equation~(\ref{eq:cc-2}) for $c_s$, and consequently the same causality conditions (\ref{eq:cc-3}), (\ref{eq:abc-1}) we found earlier from the linearized analysis in momentum space (recall that the formulas in Sec.~\ref{sec:linear-analysis} are written in a thermodynamic frame with $\theta_2=\theta_1$). In other words, the real-space analysis of hyperbolicity and causality gives the same constraints on the transport parameters as the linearized analysis of Sec.~\ref{sec:linear-analysis}.

As in Refs.~\cite{Bemfica:2017wps, Bemfica:2019knx}, it is straightforward to couple Eq.~(\ref{eq:UPDE}) to Einstein's equations. The latter have no terms with second derivatives of $\beta^\alpha$ or $\mu/T$. Thus, extending the variables to $U^A = (\beta^\alpha, \mu/T, g_{\mu\nu})$, the part of the determinant (\ref{eq:Mxi}) describing the metric degrees of freedom decouples, and the conditions of  Sec.~\ref{sec:linear-analysis} give rise to causal Navier-Stokes equations coupled to dynamical gravity.

\section{Conclusions}
We have proposed a class of stable and causal frames for relativistic hydrodynamics with conservation laws given by Eqs.~(\ref{eq:TJcons}), and whose constitutive relations contain up to one derivative of the hydrodynamic variables $T$, $u^\alpha$, and $\mu$. The stable and causal frames generalize the frames proposed in BDNK~\cite{Bemfica:2017wps, Kovtun:2019hdm, Bemfica:2019knx} to fluids with a global $U(1)$ charge such as the baryon number. A choice of frame ultimately amounts to a convention specifying how one chooses to fix the arbitrariness of defining $T$, $u^\alpha$, and $\mu$ beyond the perfect-fluid approximation. In a causal frame, the Navier-Stokes equations are causal both in flat and in curved space. 

The entropy production for hydrodynamics in the stable and causal frames is exactly the same as the entropy production in the classic Eckart or Landau-Lifshitz frames~\cite{LL6}. This is because in first-order hydrodynamics the divergence of the entropy current has to be evaluated on-shell (when the hydrodynamic equations are satisfied), and in the derivative expansion. Just like in Ref.~\cite{Kovtun:2019hdm}, the divergence of the entropy current on-shell and to first order in derivatives is the same as in~\cite{LL6}, and is only determined by $\eta$, $\zeta$, and $\sigma$.

The procedure that gives rise to the stable and causal Navier-Stokes equations is quite straightforward, and embodies the spirit of effective field theory. In the standard quantum field theory, the effective description is constructed by writing down the action in terms of all possible operators consistent with the symmetry, up to a given dimension, and then constraining the coefficients of these operators based on the stability of the vacuum and unitarity. Similarly, in hydrodynamics we write down all possible terms in the constitutive relations up to a given derivative order, and then constrain the coefficients of these terms based on the stability of equilibrium and causality. 

The most general one-derivative constitutive relations are given by Eqs.~(\ref{eq:TJ1}) and (\ref{eq:EPQTNJ1}). Using a frame in which $\varepsilon_3 = \pi_3 = \theta_3 = 0$, the hyperbolicity and causality of the equations are easily demonstrated, provided the remaining coefficients obey the inequalities discussed in Sec.~\ref{sec:linear-analysis}. One can further choose a frame in which the causality constraints look exactly like the constraints in uncharged fluids, plus a lower bound on the coefficient $\nu_3$, see Eqs.~(\ref{eq:cc-sound-1})--(\ref{eq:cc-sound-5}).

We have not performed an exhaustive analysis of stability. While the stability of the gaps requires the inequalities~(\ref{eq:stabgap-1}) to hold, a full study of stability would require working with a specific equation of state. Performing such a study would be straightforward. In conformal theories, it appears that the conditions of stability and causality at large $k$ combined with the stability at small $k$ also ensure stability at all $k$, though at the moment we do not have proof of that. We hope that our observations will stimulate further work on stable and causal relativistic Navier-Stokes equations, including their applications to heavy-ion physics and astrophysics.

\subsection*{Acknowledgments}
We would like to thank Jorge Noronha for helpful correspondence. This work was supported in part by the NSERC of Canada.

\appendix
\section{Connection to the Landau-Lifshitz frame}
In order to set up the initial value problem for the hydrodynamic equations (\ref{eq:TJcons}) in the general frame%
\footnote{
Again, we use the unfortunate but a well-established term ``frame'' to describe conventions of how one chooses to define hydrodynamic variables beyond perfect fluids.
}
 (\ref{eq:EPQTNJ1}), one needs to specify the initial values of the hydrodynamic variables $T$, $u^\alpha$, $\mu$, as well as their initial time derivatives. Setting the mathematical details aside, one is faced with a physics question: for a given non-equilibrium initial state, one needs to know how to determine the initial $T$, $u^\alpha$, $\mu$ in that state. This is non-trivial, given that the very notion of the non-equilibrium hydrodynamic variables is frame-dependent. However, if one has access to $T^{\mu\nu}$ and $J^\mu$ at early times, the question can be answered within the derivative expansion of hydrodynamics. In order to do so, one can first find $T$, $u^\alpha$, $\mu$ in the Landau-Lifshitz frame (which is relatively straightforward), and then transform those expressions to the general frame. We outline this procedure below.

In the Landau-Lifshitz frame, the hydrodynamic variables $T_L$, $u^\alpha_L$, $\mu_L$ are related to the energy-momentum tensor $T^{\mu\nu}$ and the current $J^\mu$ in the following way:
\begin{align}
\label{eq:TL}
  & T^{\alpha\beta} u_{L \beta} = -\epsilon(T_L,\mu_L) u^\alpha_L\,,\\
\label{eq:JL}
  & J^\alpha u_{L\alpha} = -n(T_L, \mu_L)\,,
\end{align}
where the functions $\epsilon$ and $n$ in the right-hand side are given by the equilibrium equation of state. Recall that $T^{\alpha\beta}$ and $J^\alpha$ are the expectation values of the corresponding microscopic operators in a given non-equilibrium state, and as such, do not depend on one's choice of convention/frame. Thus, for a given $T^{\alpha\beta}(x)$, one can in principle find $u^\alpha_L(x)$ as the timelike eigenvector of $T^{\alpha\beta}$, normalized such that $(u_L)^2 = -1$. Then, Eqs.~(\ref{eq:TL}), (\ref{eq:JL}) give $\epsilon(T_L(x),\mu_L(x))$ and $n(T_L(x),\mu_L(x))$, at each $x$. As the functions $\epsilon$ and $n$ are known from the equation of state, one can in principle reconstruct $T_L(x)$ and $\mu_L(x)$.
 
Consider now the energy-momentum tensor and the current (\ref{eq:TJ1}). Following Ref.~\cite{Kovtun:2012rj}, we write ${\cal E} = \epsilon(T,\mu) + f_{\cal E}$, ${\cal N} = n(T,\mu) + f_{\cal N}$, where $f_{\cal E}$ and $f_{\cal N}$ are the derivative corrections,
\begin{align}
\label{eq:fE}
  &  f_{\cal E} = \varepsilon_1 \dot T/T + \varepsilon_2 \nabla_{\!\lambda} u^\lambda + \varepsilon_3 u^\lambda \partial_\lambda(\mu/T) + O(\partial^2)\,,\\
\label{eq:fN}
  & f_{\cal N} = \nu_1 \dot T/T + \nu_2 \nabla_{\!\lambda} u^\lambda + \nu_3 u^\lambda \partial_\lambda(\mu/T) + O(\partial^2)\,,
\end{align}
as well as 
\begin{align}
   {\cal Q}^\mu & = \theta_1  \dot u^\mu + \theta_2/T \,\Delta^{\mu\lambda} \partial_\lambda T + \theta_3 \Delta^{\mu\lambda} \partial_\lambda (\mu/T) +  O(\partial^2)\,.
\end{align}
The hydrodynamic variables $T$, $u^\alpha$, $\mu$ are related to $T_L$, $u^\alpha_L$, $\mu_L$ by
\begin{align}
\label{eq:dddL1}
  T = T_L - \delta T, \ \ \ \ u^\alpha = u^\alpha_L - \delta u^\alpha, \ \ \ \ \mu = \mu_L - \delta \mu\,,
\end{align}
where $\delta T$, $\delta u^\alpha$, $\delta\mu$ are $O(\partial)$. Demanding that $T^{\alpha\beta}$ and $J^\alpha$ are frame-independent, one finds~\cite{Kovtun:2012rj}
\begin{align}
\label{eq:dddL2}
  \delta u^\alpha = \frac{{\cal Q}^\alpha}{\epsilon+p}\,,\ \ \ \ 
  \delta T = \frac{f_{\cal E} \frac{\partial n}{\partial\mu} - f_{\cal N} \frac{\partial \epsilon}{\partial\mu}}{ \frac{\partial\epsilon}{\partial T} \frac{\partial n}{\partial\mu} - \frac{\partial n}{\partial T} \frac{\partial\epsilon}{\partial\mu} }\,,\ \ \ \ 
  \delta \mu = \frac{ -f_{\cal E} \frac{\partial n}{\partial T} + f_{\cal N} \frac{\partial \epsilon}{\partial T}}{ \frac{\partial\epsilon}{\partial T} \frac{\partial n}{\partial\mu} - \frac{\partial n}{\partial T} \frac{\partial\epsilon}{\partial\mu} }\,.
\end{align}
The right-hand side can be evaluated in any frame, as the difference only appears at $O(\partial^2)$. For example, applying Eqs.~(\ref{eq:fE}) -- (\ref{eq:dddL2}) to uncharged fluids, we find
\begin{align}
\label{eq:uL2}
   & u^\alpha = u^\alpha_L - \frac{1}{\epsilon(T_L)+p(T_L)} \left( \theta_1(T_L) \dot u^\alpha_L + \theta_2(T_L)/T_L \,\Delta^{\alpha\lambda}_L \partial_\lambda T_L \right) + O(\partial^2)\,,\\[5pt]
\label{eq:TL2}
   & T = T_L - \frac{1}{\epsilon'(T_L)} \left( \varepsilon_1(T_L) \dot T_L/T_L + \varepsilon_2(T_L) \nabla_{\!\lambda} u^\lambda_L \right) + O(\partial^2)\,.
\end{align}
One can use the above expressions in order to find $T$, $u^\alpha$, $\mu$ in the general frame, if $T^{\mu\nu}$ and $J^\mu$ happen to be known. Given only $T^{\mu\nu}(\x,t{=}t_0)$ and $J^\mu(\x,t{=}t_0)$, one can not tell whether these single-time values correspond to a physical state that is describable by hydrodynamics. However, if one has access to $T^{\mu\nu}(\x,t)$ and $J^\mu(\x,t)$ for a range of times around $t_0$, then the time derivatives of the Landau-Lifshitz variables can be evaluated, the importance of the derivative corrections at $t=t_0$ can be estimated, and relations such as (\ref{eq:uL2}), (\ref{eq:TL2}) can be used in order to find the general-frame $T$, $u^\alpha$, $\mu$ in that state, within the derivative expansion. See also Ref.~\cite{Bemfica:2017wps}, sec.~VII C for related comments.

\bibliographystyle{JHEP}
\bibliography{hydro-general-biblio}

\end{document}